\documentclass[superscriptaddress,showpacs,preprintnumbers,amsmath,amssymb,
nofootinbib,twocolumn,aps,prd,10pt]{revtex4-1}
\usepackage{amssymb,amsmath}
\usepackage{epsfig}
\usepackage{xcolor}
\usepackage[utf8]{inputenc}
\usepackage{stmaryrd}
\usepackage{mathrsfs}
\usepackage{mathalfa}
\usepackage[normalem]{ulem}

\newcommand{\be}{\begin{equation}}
\newcommand{\ee}{\end{equation}}
\newcommand{\ba}{\begin{eqnarray}}
\newcommand{\ea}{\end{eqnarray}}
\newcommand{\beg}{\begin{gather*}}
\newcommand{\eng}{\end{gather*}}
\newcommand{\hh}{,\hspace{0.5cm}}
\newcommand{\hhh}{,\hspace{0.2cm}}

\newcommand{\lap}{\triangle}
\newcommand{\n}[1]{\label{#1}}

\newcommand{\ts}[1]{{\boldsymbol{#1}}}

\newcommand\uprule{\rule{0mm}{2ex}}

\def\XXint#1#2#3{{\setbox0=\hbox{$#1{#2#3}{\int}$ }
\vcenter{\hbox{$#2#3$ }}\kern-.6\wd0}}

\usepackage[makeroom]{cancel}
\usepackage[caption=false]{subfig}
\usepackage[colorlinks=true,
            citecolor=green,
            linkcolor=red,
            filecolor=cyan,
            urlcolor=magenta,
            backref=false]{hyperref}

\newcommand{\dd}{\mbox{d}}

\begin{document}

\title{Ultrarelativistic charged and magnetized objects\\in non-local ghost-free electrodynamics}

\author{Jens Boos}
\email{jboos@wm.edu}
\affiliation{Theoretical Physics Institute, University of Alberta, Edmonton, Alberta, Canada T6G 2E1}
\affiliation{High Energy Theory Group, Department of Physics, William \& Mary, Williamsburg, VA 23187-8795, United States}
\author{Valeri P. Frolov}
\email{vfrolov@ualberta.ca}
\affiliation{Theoretical Physics Institute, University of Alberta, Edmonton, Alberta, Canada T6G 2E1}
\author{Jose Pinedo Soto}
\email{pinedoso@ualberta.ca}
\affiliation{Theoretical Physics Institute, University of Alberta, Edmonton, Alberta, Canada T6G 2E1}

\date{\today}

\begin{abstract}
We study a non-local ghost-free Lorentz invariant modification of the Maxwell equations in four- and higher-dimensional flat spacetimes. We construct solutions of these equations for stationary charged and magnetized objects and use them to find the field created by such objects moving with the speed of light.
\end{abstract}


\maketitle

\section{Introduction}

An electric field of a point charge is spherically symmetric and its equipotential surfaces are spheres. If such a charge moves with a constant velocity with respect to an observer then its equipotential surfaces are squeezed and take on an elliptic form, which is a direct result of relativistic Lorentz contraction. In the limit when the velocity of the charge tends to the speed of light, this squeezing is so strong that the field of the charge  is practically confined to a null plane and becomes similar to a plane wave (see Ref.~\cite{Bonnor:1969rb} and references therein). A similar effect is well known in gravity: the Aichelburg--Sexl solution \cite{Aichelburg:1970dh} of the Einstein equations for an ultrarelativistic gravitating object is of the form of a so-called $pp$-wave. Bonnor \cite{Bonnor:1970sb} obtained a similar solution for a spinning gravitating object. These  results were later generalized to higher dimensions \cite{Frolov:2005in,Frolov:2005zq,Frolov:2005ja}. These so-called gyraton metrics are exact solutions of the higher-dimensional Einstein equations and they belong to the class of Kundt metrics \cite{Stephani:2003tm}. Linearized versions of the gyraton metrics can be obtained by boosting a stationary solution of the linearized Einstein equations for a spinning massive object \cite{Frolov:1418196}. In order to obtain a physically meaningful result one needs to keep the energy fixed when taking the speed of light limit, which has been dubbed ``Penrose limit'' \cite{Penrose1976}.

Recently, there has been substantial activity devoted to the study of non-local generalizations of General Relativity. The main motivation of this study is an attempt to solve  long standing problems of General Relativity: cosmological and black hole singularities. The proposed modification preserves the local Lorentz invariance of the theory. At the linear level the standard $\Box$ operator is changed to ${\cal D}=f(\Box)\Box$, where a non-local form factor $f(z)$ is chosen such that it does not vanish in the complex plane of $z$, and hence it has a unique inverse. As a result, no new unphysical degrees of freedom are present (at least at tree level). For this reason, such non-local theories are sometimes refereed to as ``ghost-free'' \cite{Biswas:2011ar,Biswas:2013cha}. Linearized solutions of the non-local ghost-free equations for stationary objects were derived and discussed in many publications (see Refs.~\cite{Boos:2018bxf,Buoninfante:2018stt} and references therein). One of the main results is that the field produced by localized objects is regularized and finite at the position of the source. The gravitational field of four-dimensional and higher-dimensional ultrarelativistic massive and spinning objects in linearized non-local gravity was found and discussed in a recent work \cite{Boos:2020ccj}; see also Ref.~\cite{Dengiz:2020xbu}. In particular, in this paper we proved a dimensional reduction formula for static Green functions: a static Green function ${\cal G}_d$ in a $(d+1)$-dimensional spacetime has the following remarkable dimensional reduction property in the Penrose limit: ${\cal G}_{d+1}\to {\cal G}_d \delta(u)$, where $u$ is a retarded null coordinate \cite{Boos:2020ccj}.

In the present paper we study the properties of a non-local ghost-free modification of electrodynamics. To that end, we adopt the formalism developed in \cite{Boos:2020ccj} for the study of the properties of the electromagnetic field in this theory created by ultrarelativistic charged and magnetized objects. We begin by collecting some known results for standard local Maxwell theory in Sec.~II in order to fix the notation and determine the consistency conditions for the scaling of parameters in the Penrose limit. We then consider a fairly general gauge-invariant non-local modification of the Maxwell equations in Sec.~III. Their solutions for point-like and extended charged and magnetized ultrarelativistic objects are given in Secs.~IV and V, respectively. These results are obtained in any number of spacetime dimensions $D\ge 4$. Last, in Sec.~VI we discuss the obtained results.

\section{Ultrarelativistic charged and magnetized objects in local electrodynamics}

Our goal is to obtain the field of ultrarelativistic charged and magnetized objects in a non-local ghost-free modification of Maxwell theory. However, it is instructive to discuss first a similar problem in the standard \emph{local} Maxwell theory. Namely, let us consider electrically charged and magnetized pencil-like objects. In both cases the transverse size of the pencil is infinitely small. We denote the length of the pencil by $\bar{L}$. We moreover consider an inertial frame $\bar{S}$ in which this pencil is at rest and denote the coordinates in this frame by $\bar{X}^{\mu}$. We specify these coordinates such that one of the spatial axes is directed along the linear extension of the pencil, and we denote this coordinate by $\bar{\xi}$, while the coordinates in the directions orthogonal to the pencil are labeled $\vec{x}_{\perp}$. Thus we have
\begin{align}
\bar{X}^{\mu}=(\bar{t},\bar{\xi},\vec{x}_{\perp})\, .
\end{align}
We choose the origin of the coordinate system such that the end points of the pencil are located at $\bar{\xi}=\pm \bar{L}/2$. In what follows, we shall boost the pencil in the $\bar{\xi}$-direction.

In this section we discuss two types of pencils. One is a uniformly charged pencil with a total electric charge $\bar{q}$, and the second type corresponds to a uniformly magnetized pencil with a total magnetic moment $\bar{m}$. To distinguish these cases we refer to them as (i) $q$-pencil and (ii) $m$-pencil, respectively.

\subsection{Field of a $q$-pencil in its rest frame}

The Maxwell equations are
\begin{align}
\begin{split}
\nabla{}_\nu F^{\nu\mu} &\equiv {1\over \sqrt{-g}} \partial_{\nu} \left( \sqrt{-g}F^{\nu\mu}\right)= j^{\mu} \, , \n{MAX}\\
 F_{\mu\nu} &=\partial_\mu A_\nu - \partial_\nu A{}_\mu \, .
\end{split}
\end{align}
Here and in what follows we shall employ Heaviside units\footnote{For more information on unit systems in electrodynamics we refer to Jackson \cite{Jackson:1999} as well as Hehl \& Obukhov \cite{Hehl:2003}.} and set the speed of light to unity, $c \equiv 1$. Let us emphasize that the background metric of the spacetime is flat, however in what follows we shall also employ non-Cartesian coordinates such that the covariant form of the Maxwell equations is very useful.

We start with a case of a $q$-pencil and assume that its charge density distribution is uniform inside a fixed interval. If $\bar{q}$ is an electric charge and $\bar{L}$ the length of the pencil, then its 4-current is
\begin{align}
\bar{j}{}^\mu = \delta{}^\mu_{\bar{t}} \bar{j}^{\bar{t}} \, , \quad
\bar{j}^{\bar{t}}={\bar{q}\over \bar{L}}\delta^{(2)}(\vec{x}_{\perp})\Theta(\bar{\xi}|-\bar{L}/2,\bar{L}/2)\, .
\end{align}
Here $\Theta(x|x_-,x_+)=\theta(x-x_-)\theta(x_+ -x)$ is a step function equal to 1 in the interval $(x_-,x_+)$ and zero outside it.
In the Coulomb gauge we may choose the vector potential to be of the form\footnote{$\ts{A}$ as a differential form is invariant under Lorentz transformations. For this reason we omit the bar on any bold-faced objects here and in what follows.}
\begin{align}\n{POT_T}
\ts{A}\equiv \bar{A}_{\mu} \dd\bar{X}^{\mu}=\bar{\phi} \dd\bar{t}\, .
\end{align}
Solving the field equation \eqref{MAX} for the potential $\bar{\phi}$,
\begin{align}
\bar{\lap}\bar{\phi} = -\bar{\lambda} \, ,
\end{align}
one finds
\begin{align}\n{PP}
\bar{\phi}=\frac{\bar{q}}{4\pi\bar{L}}\int\limits_{-\bar{L}/2}^{\bar{L}/2} {\dd \bar{\xi}'\over \sqrt{\uprule (\bar{\xi}-\bar{\xi}')^2+\rho^2}}\, ,
\end{align}
where $\rho^2=(\vec{x}_{\perp})^2$. Taking the integral one obtains
\begin{align}
\bar{\phi}=\frac{\bar{q}}{4\pi\bar{L}}\ln \left( { \bar{\xi}_+ + \sqrt{\bar{\xi}_+^2+\rho^2}\over \bar{\xi}_- + \sqrt{\bar{\xi}_-^2+\rho^2}}\right)\, ,
\end{align}
where we defined $\bar{\xi}_{\pm}=\bar{\xi}\pm \bar{L}/2$ for convenience.

\subsection{Field of an $m$-pencil in its rest frame}
Let $\{\rho,\varphi\}$ be polar coordinates in the plane orthogonal to the pencil such that we can rewrite the Minkowski metric as\footnote{Recall that the $\varphi$-component does not refer to an orthonormal basis but rather the $\partial_\varphi$-vector with norm $\rho$. Care should be taken when comparing our results to the literature, where sometimes we find expressions evaluated in orthonormal frames with the unit basis vector $\hat{\varphi} = \partial_\varphi/\rho$.}
\begin{align}
\dd s^2=-\dd\bar{t}^2+\dd\bar{\xi}^2+\dd\rho^2+\rho^2 \dd\varphi^2\, .
\end{align}
To obtain the field of the $m$-pencil let us consider first the magnetic field of a solenoid with current density
\begin{align}
\begin{split}
\ts{J} &= \bar{J}_{\varphi} \dd\varphi \, ,\ \
\bar{J}_{\varphi} = {\bar{m}\over \pi\bar{L}R}\delta(\rho-R)\Theta(\bar{\xi}|-\bar{L}/2,\bar{L}/2)\, .
\end{split}
\end{align}
Here $R$ is the radius of the solenoid, $\bar{L}$ is its length measured in the frame $\bar{S}$, and $\bar{m}$ denotes the magnetic moment of the solenoid which is proportional to the magnetic flux inside of it.
Since the magnetic field is static and axially symmetric one can put $\bar{\ts{A}}\equiv \bar{A}_{\mu} \dd\bar{X}^{\mu}=\bar{A}_{\varphi} \dd\varphi$, and the potential $\bar{A}_{\varphi}$ in the limit $R\to 0$ is
\begin{equation}\label{AF}
\bar{A}_{\varphi} = \frac{\bar{m}}{4\pi \bar{L} } \left( \frac{\bar{\xi}_+}{\sqrt{\bar{\xi}_+^2+\rho^2 }} - \frac{\bar{\xi}_-}{\sqrt{\bar{\xi}_-^2+\rho^2 }}\right)\, ,
\end{equation}
where, as earlier, $\bar{\xi}_{\pm}=\bar{\xi}\pm \bar{L}/2$. For details of this calculation we refer to Appendix \ref{app}. One can also check that the expression \eqref{AF} coincides with the magnetic field of a monopole--anti-monopole pair located on the $\bar{\xi}$-axis at the points separated by distance $\bar{L}$.\pagebreak

\subsection{Penrose limit}

Consider an inertial frame $S$ moving with respect to the rest frame $\bar{S}$ along the $\bar{\xi}$-axis with the speed $\beta$. We denote the coordinates in the boosted frame $S$ as ${X}^{\mu}=(t,\xi,\vec{x}_{\perp})$. We keep the same notation for the transverse coordinates $\vec{x}_{\perp}$ in the frame $S$ as in the rest frame $\bar{S}$ since they are not affected by the boost. The Lorentz transformation relating the coordinates in these two frames is
\begin{align}
\label{eq:boost}
\overline{t} = \gamma\left( t - \beta\xi \right) \, , \quad
\overline{\xi} = \gamma\left( \xi - \beta t \right) \, .
\end{align}
Here and in what follows we denote $\gamma=(1-\beta^2)^{-1/2}$. The second relation shows that the length $L$ of the pencil measured in the frame $S$ is $L=\bar{L}/\gamma$, which is a manifestation of the Lorentz contraction effect.

We also introduce the retarded and advanced null coordinates in the $S$ frame defined as follows:
\begin{align}\n{uv}
u = \frac{t - \xi}{\sqrt{2}} \, , \quad v = \frac{t + \xi}{\sqrt{2}} \, .
\end{align}
Then \eqref{eq:boost} implies
\begin{align}
\begin{split}
\label{tx}
\overline{t} &= {\gamma\over \sqrt{2}}[(1+\beta) u+(1-\beta)v]\, ,\\
\overline{\xi} &= {\gamma\over \sqrt{2}}[-(1+\beta) u+(1-\beta)v]\, .
\end{split}
\end{align}
In the ultrarelativistic limit, $\beta \rightarrow 1$, one has
\begin{align}\n{boost}
\overline{t} \to\sqrt{2}\gamma u \, , \quad \overline{\xi} \to -\sqrt{2}\gamma u \, .
\end{align}
We are interested in the limit of the electromagnetic field of the $q$-pencil and the $m$-pencil in the regime when $\beta\to 1$ and $\gamma\to\infty$.

Consider an observer that is at rest at the origin of the moving frame $S$. The  pencil moves to the left in this frame. Its left end point reaches the observer at the moment $u=-\hat{u}$, where $\hat{u}=L/(2\sqrt{2})$ and the right end point of the pencil passes near the observer at $u=\hat{u}$. To keep the interval $\Delta u=2\hat{u}$ finite in the limit $\gamma\to\infty$ one needs to increase the initial rest-frame size of the pencil. In other words, we put $\bar{L}=\gamma L$, where $L$ is the Lorentz contracted size of the moving pencil. The electric and magnetic charges are invariant under the boost. Hence $\bar{q}=q$, while $\bar{m}=\gamma m$, where $m$ is the magnetic moment after the boost. To summarize, we need to find a limit $\gamma\to \infty$ of the electromagnetic field of moving pencils under the conditions
\begin{align}
\n{PEN}
\bar{L}=\gamma L\hh \bar{\lambda}= \lambda/\gamma\hh\bar{\mu}=\mu\, ,
\end{align}
where $L$, $\lambda={q}/L$ and $\mu=m/L$ are kept fixed. We call this procedure the \emph{Penrose limit}.

\subsubsection{$q$-pencil}

The potential 1-form $\ts{A}$, see Eq.~\eqref{POT_T}, expressed in $\{u,v,\rho,\xi\}$ coordinates takes the form
$\ts{A}=A_{u} \dd u$. By comparing these relations one finds
\begin{align}
A_u = \sqrt{2}\gamma \bar{\phi}\, .
\end{align}
In order to find the Penrose limit of $A_u$ it is convenient first to consider the transformation of $\partial_\rho \bar{\phi}$. Using \eqref{PP} one can write
\begin{equation}\label{electboost}
\frac{\partial \bar{\phi}}{\partial \rho} = -\frac{\bar{\lambda}\rho}{4\pi} \int\limits_{-\bar{L}/2}^{\bar{L}/2}
{\dd\bar{\xi}^{\prime}\over \left[ \uprule (\bar{\xi}-\bar{\xi}^{\prime})^2+ \rho^2 \right]^{3/2}}\, .
\end{equation}
Hence
\begin{align}\n{AU}
\frac{\partial A_u}{\partial \rho} = -\frac{\gamma\bar{\lambda} \rho}{4\sqrt{2}\pi} \int\limits_{-\hat{u}}^{\hat{u}}
{\gamma \dd u'\over \left[ \uprule \gamma^2 (u-u')^2+\rho^2/2\right]^{3/2}}\, .
\end{align}
To find the limit $\gamma\to\infty$ of this integral one can use the following relation \cite{Frolov:1418196}:
\begin{equation}\label{limit}
\lim_{\gamma \to \infty} \frac{\gamma}{(\gamma^2 y^2 + \rho^2)^{m/2}} = \frac{\sqrt{\pi} \Gamma(\frac{m-1}{2})}{\Gamma(\frac{m}{2})} \frac{\delta(y)}{\rho^{m-1}} \, .
\end{equation}
In order to obtain the potential $A_u$ in the limit $\gamma\to \infty$ it is sufficient to use \eqref{boost},  to apply \eqref{limit} for $m=3$ to \eqref{AU}, to employ relation \eqref{PEN} and finally to perform the integration over $u'$. One obtains the following expression:
\begin{equation}\label{finelec}
\frac{\partial A_u}{\partial \rho} = -{\lambda \over \sqrt{2}\pi\rho} \Theta(u|-\hat{u},\hat{u})\, .
\end{equation}
The potential $A_u$ can be found by the integration of \eqref{finelec} over $\rho$. The result is
\begin{align}\n{CHAR}
 A_u=-\frac{\lambda}{\sqrt{2}\pi} \ln\left(\frac{\rho}{\rho_0}\right) \Theta(u|-\hat{u},\hat{u})\, .
\end{align}
It is easy to see that the rescaling of the integration constant $\rho_0\to C\rho_0$ can be absorbed by the gauge transformation $A_{\mu}\to A_{\mu}+\partial_{\mu} \psi(u)$. This means that the constant $\rho_0$ is physically irrelevant and does not enter observables such as the field strength $\ts{F}$.

\subsubsection{$m$-pencil}

Let us now consider the $m$-pencil and its Penrose limit. To that end, using relation \eqref{boost} in combination with the scaling property \eqref{PEN} one can write the potential $A_{\varphi}$, see Eq.~\eqref{AF}, in the form
\begin{align}
\begin{split}
A_{\varphi}={\mu \over 4\pi}&\left[ {\gamma (u+\hat{u})\over \sqrt{ \gamma^2 (u+\hat{u})^2+\rho^2/2}} \right. \\
&\left.- {\gamma (u-\hat{u})\over \sqrt{ \gamma^2 (u-\hat{u})^2+\rho^2/2}}\right] \, .
\end{split}
\end{align}
Taking the limit $\gamma\to\infty$ one finds
\begin{align}\n{MAG}
A_{\varphi}={\mu \over 2\pi}\Theta(u|-\hat{u},\hat{u})\, .
\end{align}
In the conclusion of this section we shall describe some properties of the solutions \eqref{CHAR} and \eqref{MAG} for the field of charged and magnetized pencils moving with the speed of light. Let us write the flat  Minkowski metric in $\{u,v,\rho,\varphi\}$ coordinates such that
\begin{align}
\dd s^2=-2 \dd u \dd v + \dd\rho^2 +\rho^2 \dd\varphi^2\, .
\end{align}
Let us denote by $\ts{\zeta}_{(v)}=\partial_v$ and  $\ts{\zeta}_{(\varphi)}=\partial_{\varphi}$ two of its Killing vectors. Then the solutions (\ref{CHAR}) and (\ref{MAG}) obey the following symmetries:
\begin{align}
{\cal L}_{\ts{\zeta}_{(v)}}\ts{A}={\cal L}_{\ts{\zeta}_{(\varphi)}}\ts{A}=0\, .
\end{align}
In both cases the electromagnetic field vanishes outside a finite strip $(-\hat{u},\hat{u})$ of the retarded time $u$. Moreover, on may check that the invariants $F_{\mu\nu}F^{\mu\nu}$ and $F_{\mu\nu}^{*}F^{\mu\nu}$ vanish both inside and outside this interval. In this sense the corresponding fields are similar to electromagnetic plane waves.

\section{Non-local Maxwell equations}

\subsection{Action and field equations}

In what follows we present a far going generalization of the rather simple results presented in the previous section. Namely, we consider a spacetime with arbitrary number of dimensions $D\ge 4$ and we do not assume that the electric charge and magnetic moment densities are constant. We shall also obtain results valid for both higher-dimensional Maxwell theory as well as for its non-local ghost-free generalization.

Consider $D$-dimensional flat spacetime with Cartesian coordinates $X{}^\mu = (t,\ts{x})$, with $\ts{x} = (x^i)$, $i=1,\dots,d$, and $d=D-1$. The Minkowski metric is
\begin{align}
\dd s^2 = -\dd t^2 + \sum\limits_{i=1}^d (\dd x^i)^2 = -\dd t^2 + \dd \ts{x}^2 \, .
\end{align}
In order to derive a fairly general non-local modification of the local Maxwell equations let us consider an action for the vector field $A_{\mu}$ of the form
\begin{align}\n{SS}
S[A_{\mu}]= \frac12 \int \dd^D X A_{\mu} {\cal O}^{\mu\nu} A_{\nu}\, ,
\end{align}
where ${\cal O}^{\mu\nu}$ is an arbitrary symmetric tensor function of the derivatives. It is easy to check that in a general case it has the form
\begin{align} \n{OO}
 {\cal O}^{\mu\nu}=h(\Box)g^{\mu\nu}+f(\Box)\nabla^{\mu}\nabla^{\nu}\, .
\end{align}
In Cartesian coordinates one has $g_{\mu\nu}=\eta_{\mu\nu}$ and $\nabla_{\mu}=\partial/\partial X^{\mu}$, but it is convenient to work with a covariant form of the action. One only needs to remember that in this case the operators $\nabla_{\mu}$ commute and their action on the metric vanishes, $\nabla{}_\rho g{}_{\mu\nu} = 0$. Let us also emphasize that we shall use the action solely for deriving the field equations for $A{}_\mu$ and for that reason one may integrate by parts without considering the contribution of the boundary terms.

Let us demand now that the action is invariant under the U(1) gauge transformation $A_{\mu}\to \hat{A}_{\mu}=A_{\mu}+\nabla_{\mu}\psi $, where $\psi$ is an arbitrary function of $X^{\mu}$. One finds
\begin{align}
\delta_\psi S &\equiv S[\hat{A}_\mu] -S[{A}_\mu]= \int \dd^D X \, J \, , \\
J &= -\psi \left[ h(\Box)+f(\Box)\Box \right]\nabla^\mu A_\mu \nonumber \\
&\hspace{11pt} +\frac{\psi}{2} \Box \left[ h(\Box)+f(\Box)\Box \right] \psi \, . \n{var}
\end{align}
Let us consider first the second term in the expression for $J$, which is quadratic in the gauge function $\psi$. Since this function is arbitrary, the last term vanishes only if
\begin{align} \n{ca}
h(\Box)=-f(\Box)\Box \, .
\end{align}
However, under this condition the first term on the right-hand side of \eqref{var} which is linear in $\psi$ vanishes as well. Hence the condition \eqref{ca} guarantees that the action \eqref{SS}-\eqref{OO} is gauge invariant
 taking the form
\begin{align}\n{SSS}
S[A_{\mu}]=\frac12\int \dd^D X A_{\mu} f(\Box) [{g}^{\mu\nu}\Box - \nabla^{\mu}\nabla^{\nu}] A_{\nu}\,  .
\end{align}
We add to this action a term describing an interaction with a conserved external current $j^{\mu}$, and after integration by parts write it in the form
\begin{align}
\begin{split}
S[A_\mu] &= \int\dd^D X \left[ \frac14 F_{\mu\nu} f(\Box) F{}^{\mu\nu} - j^\mu A_\mu \right] \, , \\
F_{\mu\nu} &= \partial_\mu A_\nu - \partial_\nu A{}_\mu \, .
\end{split}
\end{align}
The operator $f(\Box)$ in this action is called a \emph{form factor}, and its precise form specifies a non-local model. We assume that the corresponding function  $f(z)$ viewed as a function of the complex variable $z$ has no poles in the complex plane. In this case the operator $f(\Box)$ has an inverse and no new unphysical degrees of freedom (ghosts) emerge. We also assume that $f(0)=1$, which guarantees that the residue of the pole of $zf(z)$ at $z=1$ is 1, such that the theory correctly reproduces the properties of the corresponding local theory in the infrared regime. This class of theories is often referred to as ``ghost-free.'' Quite often the form factor $f$ is chosen to be of the form
\begin{align}
f(\Box) = \exp\left[ (-\ell^2\Box)^N \right] \, , \quad \ell > 0 \, .
\end{align}
Here $N$ is a positive integer number and $\ell > 0$ is a characteristic length scale of non-locality. We call non-local models with these form factors $\mathrm{GF_N}$ theories. In the limiting case of $\ell\rightarrow 0$ one recovers the local theory since $f(0) = 1$.

The non-local field equations take the form
\begin{align}
f(\Box) \partial_\mu F{}^{\mu\nu} &= j{}^\nu \hh
\partial_{[\rho} F_{\mu\nu]} = 0 \, .
\end{align}
The second equation implies that locally $F_{\mu\nu} = \partial_\mu A{}_\nu - \partial_\nu A{}_\mu$. Inserting this into the first equation one finds
\begin{align}
f(\Box) \left( \Box A{}^\nu - \partial^\nu \partial_\mu A{}^\mu \right) = j{}^\nu \, .
\end{align}
We may now exploit the U(1) gauge invariance in $A_\mu$ to fix the gauge to the convenient choice (``Lorenz gauge'') $\partial_\mu A{}^\mu = 0$ which implies
\begin{align}
f(\Box) \Box A_\mu \overset{*}{=} j{}_\mu \, .
\end{align}
For the remainder of this paper we shall work exclusively in the Lorenz gauge.

\subsection{Green functions}

Our first step consists of finding solutions that describe the electromagnetic field of static sources. Since the field does not depend on time one may substitute the $\Box$-operator by the Laplace operator
\begin{align}
\lap = \nabla^2 = \sum_{i=1}^d \partial_i^2\, .
\end{align}
In both the local and non-local cases a solution can be found via the corresponding Green function, which is a solution of the following differential equation:
\begin{align}
f(\lap) \lap \mathcal{G}_d(\ts{x}'-\ts{x}) = -\delta{}^{(d)}(\ts{x}'-\ts{x}) \, .
\end{align}
For a $\mathrm{GF_N}$ theory one has
\begin{align}
f(\lap) = \exp[(-\ell^2\lap)^N] \, .
\end{align}
It is possible to show that the following integral representation for the Green function $\mathcal{G}_d(r)$ is valid:
\begin{align}
\label{eq:ch2:bessel-representation-1}
\mathcal{G}_d(r) = \frac{1}{(2\pi)^{d/2}r^{d-2}} \int\limits_0^\infty \dd\zeta \zeta^{\frac{d-4}{2}} e^{-(\zeta\ell/r)^{2N}} J_{\frac{d}{2}-1}(\zeta) \, .
\end{align}
Here $d \ge 3$, and we abbreviated $r \equiv |\ts{x}'-\ts{x}|$ for convenience.
There is also a recursive formula relating these Green functions in the spaces of different dimensions,
\begin{align}
\label{eq:ch2:recursion}
\mathcal{G}_{d+2}(r) = -\frac{1}{2\pi r} \frac{\partial \mathcal{G}_d(r)}{\partial r} \, .
\end{align}
These results and their derivation can be found in \cite{Frolov:2015usa}.

The first of these relations allows one to find the Green functions in an explicit form for some special non-local models. If such a Green function is found for $d=3$ and $4$, the higher-dimensional Green functions can be found by simple differentiation via Eq.~\eqref{eq:ch2:recursion}. For example, in the simplest case of $\mathrm{GF_1}$ theory, $N=1$, one has
\begin{align}
\mathcal{G}_3(r) &= \frac{1}{4\pi r}\text{erf}\left( \frac{r}{2\ell} \right) \, , \\
\mathcal{G}_4(r) &= \frac{1}{4\pi^2 r^2}\left[ 1 - e^{-r^2/(4\ell^2)} \right] \,  ,
\end{align}
where $\text{erf}(z)$ denotes the error function \cite{Olver:2010}. In the limit $\ell\rightarrow 0$ one recovers the well-known local expressions. Moreover, all non-local ghost-free Green functions are manifestly regular at $r=0$.

\section{Point-like sources}

Both local Maxwell theory and its non-local modification are linear theories. This means that it is sufficient to find a solution for a point-like source. The field of an extended object can be obtained by integrating such a solution over the volume occupied by the object with a proper weight representing the charge and magnetic moment density distributions.

\subsection{Stationary point particle}

Let us consider the conserved external current
\begin{align}
j{}^\mu = \delta{}^\mu_t q \delta{}^{(d)}(\ts{x}) + \delta{}^\mu_i M{}^{ik} \partial_k \delta{}^{(d)}(\ts{x}) \, .
\end{align}
Here $i,k=1,\ldots,d$, $q$ is the charge of the point particle and $M_{ik} = -M_{ki}$ is a constant, antisymmetric tensor that parametrizes the particle's intrinsic magnetic moment. Writing the electromagnetic potential as
\begin{align}
A_\mu(\ts{x}) = \delta{}_\mu^t \varphi(\ts{x}) + \delta{}_\mu^i A_i (\ts{x})
\end{align}
the equations of motion take the form
\begin{align}
f(\lap)\lap \varphi &= -q\delta{}^{(d)}(\ts{x}) \, , \\
f(\lap)\lap A_i &= M{}_i{}^k \partial_k \delta{}^{(d)}(\ts{x}) \, .
\end{align}
These equations are solved by
\begin{align}
\varphi(\ts{x}) = q \, \mathcal{G}_d(\ts{x}) \, , \quad
A_i(\ts{x}) = -M{}_i{}^k \partial_k \mathcal{G}_d(\ts{x}) \, .
\end{align}
The potential 1-form $\ts{A} = A{}_\mu\dd X{}^\mu$ then takes the form
\begin{align}
\begin{split}\n{APOT}
A_\mu \dd X{}^\mu &= \varphi \dd t + A_i \dd x{}^i \\
&= q \, \mathcal{G}^N_d(r) \dd t - M{}_i{}^k \partial_k \mathcal{G}_d(\ts{x}) \dd x^i  \, .
\end{split}
\end{align}
Using the relation
\begin{align}
\partial_i \mathcal{G}_d(\ts{x}) = \frac{x_i}{r} \partial_r \mathcal{G}_d(\ts{x}) = -2\pi x_i \mathcal{G}_{d+2}(\ts{x})
\end{align}
we may also write
\begin{align}
\begin{split}\n{APOT2}
A_\mu \dd X{}^\mu = q \, \mathcal{G}^N_d(r) \dd t + 2\pi M{}_i{}^k x_k \mathcal{G}_d(\ts{x}) \dd x^i  \, .
\end{split}
\end{align}

\subsection{Boosting a point-like source}

As in Sec.~II we single out a boost direction and denote the coordinates as $\bar{X}{}^\mu = (\bar{t}, \bar{\xi}, \ts{x}_\perp)$ such that
\begin{align}
\dd s^2 = -\dd\bar{t}^2 + \dd\bar{\xi}^2 + \dd\ts{x}_\perp^2 \, .
\end{align}
As earlier, we denote quantities calculated in the source's rest frame $\bar{S}$ with bars. The space orthogonal to $\bar{\xi}$ is $(d-1)$-dimensionsal and we denote $d-1=2n+\epsilon$.

\begin{figure}[!hbt]
    \centering
    \vspace{10pt}
    \includegraphics[width=0.48\textwidth]{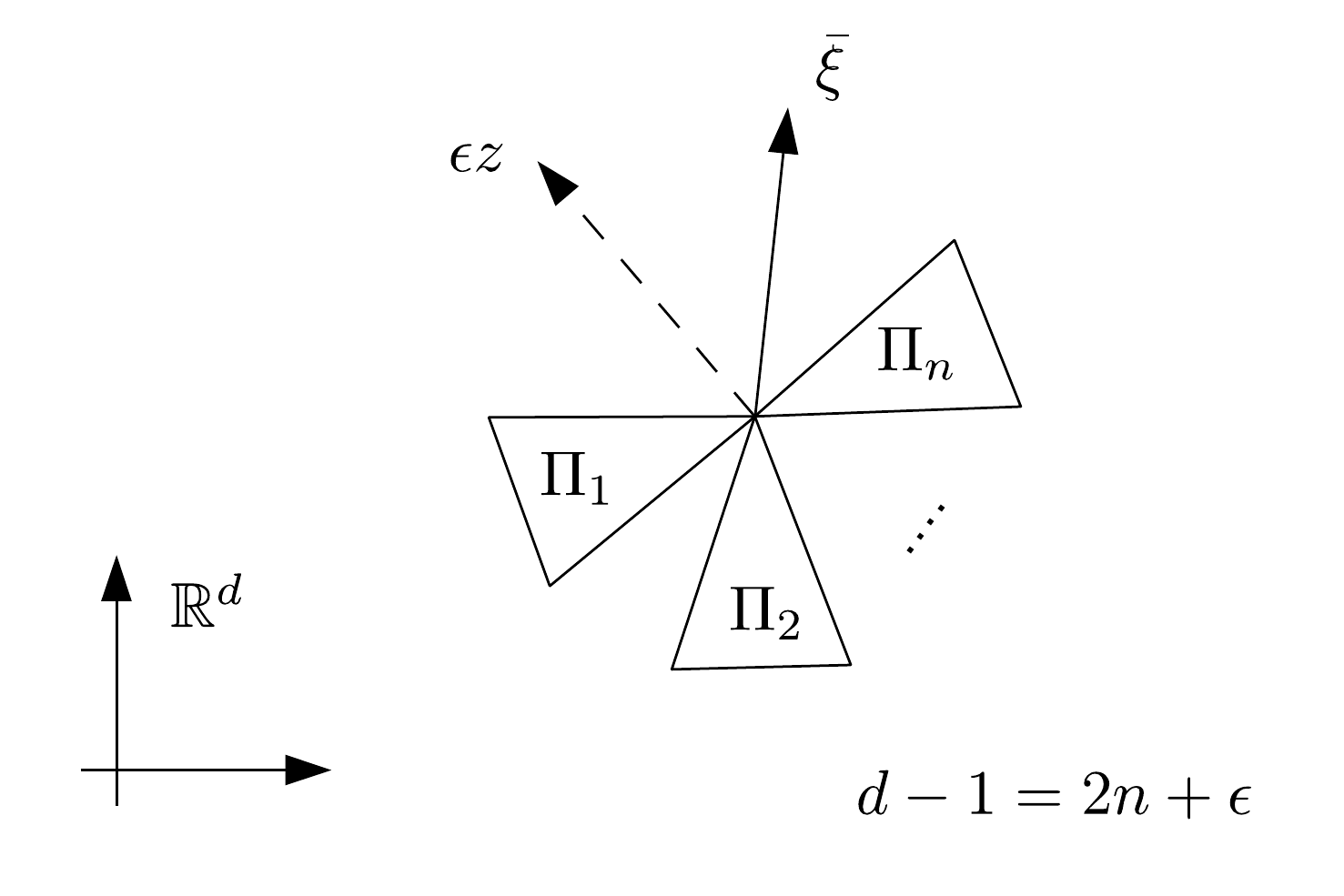}
    \caption{Darboux decomposition of $d$-dimensional space into $n$ mutually orthogonal Darboux planes $\Pi_a$ and a transverse $z$-direction if $\epsilon=1$ \cite{Boos:2020qgg}.}
    \label{fig:darboux}
\end{figure}

In that transverse space, we choose $n$ mutually orthogonal two-planes $\Pi_a$, $a=1,\ldots,n$. If $d$ is odd ($\epsilon=0$) then the $n$ two-planes span the complete transverse space. For $d$ even one has $\epsilon=1$ and in order to fully span the transverse space, besides $n$ two-planes, one needs an additional one-dimensional space. We call the corresponding coordinate $z$; see Fig.~\ref{fig:darboux} for a visualization of this decomposition. In a general case certainly there is an ambiguity in the choice of the set of two-planes. We assume that the spatial antisymmetric matrix $M_{ij}$ which enters the current has non-vanishing components only in the directions transverse to $\bar{\xi}$. In four spacetime dimensions, where $n=1$, this assumption implies that the vector of the magnetic moment generated by the current is directed along the $\bar{\xi}$-direction. The above condition imposed on $M_{ij}$ plays a similar role in higher dimensions. Let us note that by rigid rotations in the space transverse to $\bar{\xi}$ one can find coordinates $\{y_a,\hat{y}_a\}$ in which the $(d-1)$-dimensional metric is
\begin{align}
\dd l^2 = \sum_{a=1}^n (\dd y_a^2 + \dd\hat{y}_a^2) + \epsilon \dd z^2 \, ,
\end{align}
and the matrix $M_{ij}$ takes a quasi-diagonal form with antisymmetric $2\times 2$ blocks,
\begin{align}
M_{ij} =
\begin{pmatrix}
0 & \bar{m}_a \\
-\bar{m}_a & 0
\end{pmatrix} \, .
\end{align}
A plane $\Pi_a$ spanned by the coordinates $\{y_a,\hat{y}_a\}$ is called a two-eigenplane or ``Darboux plane'' of $M_{ij}$. 

In what follows it is convenient to introduce polar coordinates $\{\rho_a,\varphi_a\}$ in each of the two-planes $\Pi_a$, such that the metric takes the form
\begin{align}
\dd s^2 = -\dd\bar{t}^2 + \dd \bar{\xi}^2 + \sum_{a=1}^n (\dd\rho_a^2 +\rho_a^2 \dd\varphi_a^2)+\epsilon \dd z^2\, .
\end{align}
In these coordinates the field of the point-like source \eqref{APOT} can be expressed as
\begin{align}
\label{eq:a:pre-boost}
\bar{A}_\mu \dd \bar{X}{}^\mu &= \bar{q} \, \mathcal{G}_d(\bar{r}) \dd \bar{t} - 2\pi \mathcal{G}_{d+2}(\bar{r}) \sum\limits_{a=1}^n \bar{m}_a \rho_a^2 \dd\varphi_a \, , \nonumber \\
\bar{r}^2 &= \bar{\xi}^2 + \ts{x}_\perp^2 = \bar{\xi}^2 + \sum\limits_{a=1}^n \rho_a^2 + \epsilon z^2 \, .
\end{align}
As earlier, we consider a new boosted frame $S$ moving along the $\bar{\xi}$-axis with the speed $\beta$. The coordinate transformation relating $\{\bar{t},\bar{\xi}\}$ to the $\{t,\xi\}$ coordinates in the $S$-frame is given by the Lorentz transformation \eqref{eq:boost}. Since the transverse coordinates $\ts{x}_\perp$ are not affected by the boost the relations \eqref{uv}--\eqref{boost} are also valid for higher-dimensional cases.

We may now substitute these relations into \eqref{eq:a:pre-boost} and obtain the boosted vector potential for finite $\gamma$,
\begin{align}
\label{eq:boosted-particle}
A_\mu \dd X{}^\mu &= \sqrt{2}\bar{q}\gamma \mathcal{G}_d(\bar{r}) \dd u - 2\pi \mathcal{G}_{d+2}(\bar{r}) \sum\limits_{a=1}^n \bar{m}_a \rho_a^2 \dd\varphi_a \, , \nonumber \\
\bar{r}^2 &= 2\gamma^2 u^2 + \ts{x}_\perp^2  \, .
\end{align}
In order to obtain the Penrose limit of the solution \eqref{eq:boosted-particle} we assume that $q=\bar{q}$ and $m_a=\sqrt{2}{\bar m}_a/\gamma$. To perform the limit we shall also use the following result \cite{Boos:2020ccj}:
\begin{align}
\label{eq:master}
\lim\limits_{\gamma\rightarrow\infty} \gamma \mathcal{G}_d\left( \bar{r} \right) = \frac{1}{\sqrt{2}} \mathcal{G}_{d-1}(r_\perp)\delta(u) \, ,
\end{align}
One finally obtains
\begin{align}
\begin{split}
\label{eq:max-boosted}
A_\mu \dd X{}^\mu &= q \mathcal{G}_{d-1}(r_\perp) \delta(u) \dd u \\
&\hspace{11pt}- \pi \mathcal{G}_{d+1}(r_\perp) \delta(u) \sum\limits_{a=1}^n m_a \rho_a^2 \dd\varphi_a  \, .
\end{split}
\end{align}
The spacetime metric in $\{u,v\}$ coordinates is
\begin{align} \n{uv-metric}
\dd s^2=-2\dd u \dd v + \sum_{a=1}^n (\dd\rho_a^2 +\rho_a^2 \dd\varphi_a^2)+\epsilon \dd z^2\, .
\end{align}
For the non-local theory the potential of an ultrarelativistic point particle \eqref{eq:max-boosted} has two interesting features: first, it is regular as $r_\perp \rightarrow 0$, which is in stark contrast to the results one obtains in standard local Maxwell theory. And second, the appearance of the $\delta(u)$-factors confirms our comments in the Introduction of this paper: the electromagnetic field is indeed confined to a null plane, just as it happens in the local case. This behavior is somewhat singular and is not cured by the presence of non-locality. This is expected, however, since it is caused by the infinite Lorentz contraction in the $\bar{\xi}$-direction, and one would not expect a Lorentz invariant non-local modification to affect this mechanism. As we shall see now, for extended objects endowed with an additional Lorentz-scaling property under boosts, these $\delta(u)$ artefacts can be avoided altogether.

\section{Ultrarelativistic extended objects}

\subsection{Boosting charged and magnetized pencils}
Let us next consider the field of charged and magnetized extended objects  and their Penrose limits.  For simplicity we limit our considerations to charged and/or magnetized pencils whose transverse charge and magnetic moment densities are $\delta$-shaped, but we allow a density profile in the pre-boost $\bar{\xi}$-direction to be arbitrary functions of $\bar{\xi}$. By linearity, results for objects with a finite transverse extension can be obtained by superimposing pencil-like solutions. Let us denote these densities by $\bar{\lambda}(\bar{\xi})$ and $\bar{\mu}_a(\bar{\xi})$ for the charged and magnetized pencils, respectively. Then, the conserved external current takes the following form:
\begin{align}
j{}^\mu = \delta{}^\mu_{\bar{t}} \bar{\lambda}(\bar{\xi}) \delta{}^{(d-1)}(\ts{x}_\perp) + \delta{}^\mu_i \bar{\mu}{}^{ik}(\bar{\xi}) \partial_k \delta{}^{(d-1)}(\ts{x}) \, .
\end{align}
In what follows we shall make an additional assumption, namely that the orientation of the $\mu^{ik}$-Darboux planes does not depend on $\bar{\xi}$ and that the magnetization has no $\bar{\xi}$-component.\footnote{For more details on antisymmetric objects in higher dimensions and their Darboux decompositions we refer to Ch.~3.9 in Ref.~\cite{Boos:2020qgg}.} Then we can define the objects $\mu_a(\bar{\xi})$ for every value of $\bar{\xi}$ such that the current can be rewritten in the form
\begin{align}
j{}^\mu = \left[ \delta{}^\mu_{\bar{t}} \bar{\lambda}(\bar{\xi}) + \sum\limits_{a=1}^n \bar{\mu}_a(\bar{\xi}) \epsilon{}^{(a)\mu j}\partial_j \right] \delta{}^{(d-1)}(\ts{x}_\perp) \, ,
\end{align}
where $\epsilon{}^{(a)}{}_{ij} = -\epsilon{}^{(a)}{}_{ji}$ is the volume element in the $a$-th Darboux plane. Then, the total charge $\bar{q}$ and magnetic moment $\bar{m}_a$ of the pencil are given by the line integrals
\begin{align}
\bar{q} = \int\limits_{-\infty}^\infty \dd\bar{\xi} \, \bar{\lambda}(\bar{\xi}) \, , \quad
\bar{m}_a = \int\limits_{-\infty}^\infty \dd\bar{\xi} \, \bar{\mu}_a(\bar{\xi}) \, .
\end{align}
If the pencil has a finite length the integrals should be taken for a finite interval of $\bar{\xi}$.

The expression for $\bar{A}{}_\mu$ then takes the form
\begin{align}
\label{eq:a:pre-boost-pencil}
\bar{A}_{\bar{t}} &= \int\limits_{-\infty}^\infty \dd\bar{\xi}' \, \bar{\lambda}(\bar{\xi}') \mathcal{G}_d(\bar{r}) \, , \\
\bar{A}_{a} &= -2\pi \int\limits_{-\infty}^\infty \dd\bar{\xi}' \, \mathcal{G}_{d+2}(\bar{r}) \sum\limits_{a=1}^n \bar{\mu}_a(\bar{\xi}') \rho_a^2 \, , \\
\bar{r}^2 &= (\bar{\xi}'-\bar{\xi})^2 + \ts{x}_\perp^2 \hh
\ts{x}_\perp^2 = \sum\limits_{a=1}^n \rho_a^2 + \epsilon z^2 \, .
\end{align}
Both $\bar{\lambda}(\bar{\xi})$ and $\bar{\mu}_a(\bar{\xi})$ are one-dimensional line densities. To satisfy the scaling laws \eqref{PEN} we hence define
\begin{align}
\lambda(u) &= \lim\limits_{\gamma\rightarrow\infty} \sqrt{2}\gamma \bar{\lambda}(-\sqrt{2}\gamma u) \, , \\
\mu_a(u) &= \lim\limits_{\gamma\rightarrow\infty} \sqrt{2} \bar{\mu}_a(-\sqrt{2}\gamma u) \, .
\end{align}
For this reason, again making use of the relation \eqref{eq:master}, one finds the following expressions for the potentials for the ultrarelativistic charged and magnetized pencils:
\begin{align}
\begin{split}
\label{eq:ultrarelativistic-pencil}
A_\mu \dd X{}^\mu &= \lambda(u) \mathcal{G}^N_{d-1}(r_\perp) \dd u \, \\
&\hspace{11pt} - \pi \mathcal{G}_{d+1}^N(r_\perp) \sum\limits_{a=1}^n \mu_a(u) \rho_a^2 \dd \varphi_a \, .
\end{split}
\end{align}
Comparing these expressions with the boosted point sources \eqref{eq:max-boosted} one finds a correspondence by formally replacing $\lambda(u) \rightarrow q\delta(u)$ and $\mu_a(u) = m_a\delta(u)$. This is expected since the linear extension of the pencil in combination with the Penrose limit were stipulated in order to remove the unphysical $\delta(u)$-factors in electromagnetic potential of the boosted sources.

\subsection{Properties of solutions}
Let us now make some remarks concerning the properties of the obtained solutions within the non-local Maxwell theory. To that end, the obtained solution \eqref{eq:ultrarelativistic-pencil} can be written in the form
\begin{align} \n{KA}
\ts{A}=\sum_{a=0}^n \lambda_a (u) a_a({r}_\perp) \ts{\zeta}_a\, ,
\end{align}
where $\lambda_0(u)=\lambda(u)$ and $\lambda_{a\ge 1}(u)=\mu_a(u)$. We denoted by $\ts{\zeta}$ the following Killing vectors:
\begin{align}
\ts{\zeta}_0=\partial_v\hh \ts{\zeta}_a=\partial_{\varphi_a}\, .
\end{align}
It is easy to check that
\begin{align}
{\cal L}_{\ts{\zeta}_a} \ts{A}=0\, ,
\end{align}
where ${\cal L}_{\ts\zeta}$ is the Lie derivative in $\ts{\zeta}$ direction. These relation show that the boosted solutions have the expected symmetries, that is, no dependence on advanced time $v$ as well as rotational isometries in the $\varphi_a$-directions.

Another observation is the following. In the absence of the magnetic moments, $\mu_a=0$, both in the local and non-local case, the electromagnetic field $\ts{F}$ is null,
\begin{align}
F{}_{\mu\alpha}F{}^\alpha{}_\nu=S u_{,\mu} u_{,\nu} \, , \quad
F^2 = \frac12 F_{\mu\nu}F^{\mu\nu}=0\, .
\end{align}
In general, the presence of the magnetic field violates this property. However, the case of electrodynamics in four-dimensional spacetime is an exception. To demonstrate this, let us write the potential 1-form $\ts{A}$ in $4D$ as
\begin{align}
\ts{A}=b(u)c(\rho)\ts{\zeta}_0+B(u)C(\rho)\ts{\zeta}_{\varphi}\, ,
\end{align}
where $\ts{\zeta}_0$ and $\ts{\zeta}_{\varphi}$ are the 1-forms that are dual to their respective Killing vector. Then, calculations show that
\begin{align}\n{FFF}
F^2= B^2 \left(\rho {\dd C\over \dd\rho}+2C\right)^2\, .
\end{align}
Thus $F^2=0$ only when $C=C_0/\rho^2$. This is precisely the case for the field of the magnetized and charged ultrarelativistic pencil in four dimensions in the framework of the standard local Maxwell theory. For this theory in higher dimensions this property is violated. Let us emphasize that in the non-local theory $F^2\ne 0$ for ultrarelativisic magnetized pencils not only in the higher dimensions, but in four spacetime dimensions as well.

\section{Conclusions}
In this paper we first derived the electromagnetic field of charged and magnetized point particles and extended objects in a flat spacetime. Using Green functions of the non-local version of the Laplace operator we obtained solutions describing the electromagnetic field for electrically charged and magnetized objects which are at rest in some inertial frame. These results were obtained for local Maxwell theory and its non-local ghost-free modification in any number of spatial dimensions. We used these solutions and boosted them to the speed of light by formulating suitable scaling properties that are required to perform such a ``Penrose limit,'' and obtained finite results. After this we studied the properties of the electromagnetic field of the resulting ultrarelativistic objects. The main insight is that the electric field of the ultrarelativistic object in $d$ spatial dimensions is given in terms of the $(d-1)$-dimensional static scalar Green function, while the corresponding expression for the magnetic field contains the $(d+1)$-dimensional static scalar Green function.
This property is similar to the property of solutions for the gravitational field of ultrarelativistic objects (gyratons) in the weak-field regime of non-local gravity which was discussed in Ref.~\cite{Boos:2020ccj}.

For a static point-like source the presence of non-locality modifies its field and renders it finite and regular at the location of the source at $r=0$. The scale of non-locality $\ell$ determines the size of the region where the non-local modification of the solution is relevant. Our results for fields of the corresponding ultrarelativistic point-like objects show that similar regularization properties hold in this case as well. Namely, functions depending on the coordinates transverse to the direction motion are all rendered finite and regular. However, these solutions contain $\delta(u)$-functions which imply that the field is non-trivial only on a single null plane where the retarded time vanishes, $u=0$. This means that at least for this class of solutions, non-locality does \emph{not} bring about causality violations in the shape of faster-than-light signalling. A similar property is valid for extended (pencil-type) objects of finite ``length'' (duration in time $u$). An observer registers a non-zero field of such an object only when it passes them; before or after that event the the field is identically zero.

Let us also mention another interesting property of the considered non-local ghost-free modification of Maxwell theory: in the standard local theory in four spacetime dimensions, the invariant $F{}_{\mu\nu}F{}^{\mu\nu}$ vanishes for an ultrarelativistic charged and magnetized object. This property is violated for magnetized objects in the non-local 4D theory, which might give rise to observable consequences.

One can expect that the described effects should manifest themselves in the scattering of two ultrarelativistic charged particles. It is interesting to study whether the presence of fundamental non-locality can in principle be experimentally tested under such conditions.

\section*{Acknowledgments}
J.B.\ is grateful for a Vanier Canada Graduate Scholarship administered by the Natural Sciences and Engineering Research Council of Canada as well as for the Golden Bell Jar Graduate Scholarship in Physics by the University of Alberta, and was supported in part by the National Science Foundation under grant PHY-1819575. V.F.\ thanks the Natural Sciences and Engineering Research Council of Canada and the Killam Trust for their financial support.

\appendix

\section{Details of calculations}
\label{app}

In this appendix we give an expression for the potential of an $m$-pencil in three spatial dimensions, which is used in the main body of this paper. The easiest way to obtain an expression for the current density of an $m$-pencil is the following. Let us consider a thin cylinder of radius $R$ and length $\bar{L}$ along the $\bar{\xi}$-axis. Suppose particles with the total electric charge $Q$, which are uniformly distributed on the cylinder, rotate around its symmetry axis with the same angular velocity $\omega$. To keep the system neutral one may add particles with the opposite charge uniformly distributed on the cylinder but which do not rotate. The magnetic moment $\textbf{m}$ of a system of charges $q_a$ located at $\ts{r}_a$ and moving with velocity $\ts{v}_a$ is given by the following expression \cite{Landau:1971}:
\begin{align} \n{mmm}
\ts{m}={1\over 2} \sum_a q_a [\ts{r}_a\times \ts{v}_a]\, .
\end{align}
In cylindrical coordinates $\{\bar{\xi},\rho,\varphi\}$ the magnetic moment of the rotating charged cylinder then takes the form
\begin{align}\n{m2}
\ts{m}=(\bar{m},0,0)\hh
\bar{m}={1\over 2} Q R^2 \omega\, .
\end{align}
The current density of a system of charged particles is
\begin{align}
\bar{\ts{J}}=\sum_a q_a \ts{v}_a\delta{}^{(d)}(\ts{r}-\ts{r}_a)\,  .
\end{align}
For the rotating cylinder one finds
\begin{align}
\begin{split}
\bar{\ts{J}} &= (0,0,\bar{J}^{\varphi} ) \, , \\
\bar{J}^{\varphi} &= {Q\omega \over 2\pi R \bar{L} } \delta(\rho-R) \Theta\left(\bar{\xi}|-\bar{L}/2,\bar{L}/2\right) \, .
\end{split}
\end{align}
The field equations \eqref{MAX} imply the following equation for the potential $\bar{A}_{\varphi}$:
\begin{align} \n{A1}
\rho \partial_{\rho} \left( {1\over \rho} \partial_{\rho}\bar{A}_{\varphi}\right)
+\partial_{\bar{\xi}}^2\bar{A}_{\varphi} = \bar{J}_{\varphi}\, .
\end{align}
Denoting $\partial_{\rho}\bar{A}_{\varphi}=\rho Z$, Eq.~\eqref{A1} gives
\begin{align}
 {1\over \rho} \partial_{\rho} \left( \rho\partial_{\rho} Z \right)
+\partial_{\bar{\xi}}^2 Z= j\, ,
\end{align}
where $j(\rho)= {1\over \rho}\partial_{\rho} \bar{J}_\varphi$. The left-hand side of this equation is nothing but the flat three-dimensional Laplace operator in cylindrical coordinates applied to the scalar function $Z(\rho,\bar{\xi})$. Using the Green function of this operator, expressed in cylindrical coordinates, one then obtains
\begin{align}
Z(\rho,\bar{\xi}) &= \int\limits_{-\bar{L}/2}^{\bar{L}/2} \dd\bar{\xi}'\int\limits_0^{2\pi} \rho' \dd\varphi' P\hhh
P =-\int\limits_0^{\infty} \dd\rho' {j(\rho')\over 4\pi r}\, ,\n{ZZ}\\
r &= (\rho^2+\rho'^2-2\rho\rho' \cos\varphi' +z^2)^{1/2}\, .\n{RR}
\end{align}
Here we abbreviated $z=\bar{\xi}-\bar{\xi}'$. The integral for $P$ can be evaluated with the following result:
\begin{align}
P={Q\omega R\over 8\pi^2 \bar{L}} {\partial \over \partial R} \left({1\over r}\right)\, ,
\end{align}
where $r$ is given by \eqref{RR} subject to the substitution $\rho'=R$ in this expression. Using definition \eqref{m2} of the magnetic moment one can write
\begin{align}
Z(\rho,\bar{\xi})={\bar{m}\over 4\pi^2 R \bar{L}}\int\limits_{-\bar{L}/2}^{\bar{L}/2} \dd\bar{\xi}'{\partial I\over \partial R} \, .
\end{align}
Here we introduced the shorthand notation
\begin{align}
I=\int\limits_0^{2\pi} {\dd \varphi\over r}={4 K\left( {2\sqrt{ R\rho}\over \sqrt{(\rho+R)^2+z^2}}\right)\over  \sqrt{(\rho+R)^2+z^2}}\, ,
\end{align}
where $K$ is the complete elliptic integral of second type. Expanding for small values of its argument one  finds
\begin{align} \n{FF}
 S=\lim_{R\to 0}{1\over R} \partial_{R} I=\pi {\rho^2-2z^2\over (\rho^2+z^2)^{5/2}}\, .
\end{align}
Using these results and restoring $\bar{A}_{\varphi}$ one obtains
\begin{align}
\bar{A}_{\varphi}={\bar{m}\over 4\pi \bar{L}} \int\limits_{-\bar{L}/2}^{\bar{L}/2} \dd\bar{\xi}'\int\limits_0^{\rho} \dd\rho' \rho' \, S \, .
\end{align}
Performing the integration over $\rho'$ and $\bar{\xi}'$ finally yields
\begin{equation}\label{sole}
\bar{A}_{\varphi} = \frac{ \bar{m}}{ 4\pi\bar{L} } \left( \frac{\bar{\xi}_+}{\sqrt{\rho^2 +\bar{\xi}_+^2}} - \frac{\bar{\xi}_-}{\sqrt{\rho^2 +\bar{\xi}_-^2}}\right)\, ,
\end{equation}
where we abbreviated $\bar{\xi}_{\pm}=\bar{\xi}\pm \bar{L}/2$. Let us mention that this expression for the field of a magnetized infinitely thin pencil can also be obtained by using the expression for the potential of a magnetized solenoid of finite radius $R$, which can be found in Jackson's book \cite{Jackson:1999}.

\bibliography{Ghost_references,GYRATON,electrodynamics}{}

\end{document}